\documentclass[a4paper,11pt]{article}
\usepackage[ams]{}
\usepackage{amssymb}
\usepackage{graphicx}
\usepackage[theorem]{}

\begin{document}
\title{Jordan Algebras and Extremal Black Holes}
\author{Michael Rios\footnote{email: mrios4@calstatela.edu}\\\\\emph{California State University, Los Angeles}\\\emph{Physics Graduate Program}
\\\emph{5151 State University Drive}\\\emph{Los Angeles, CA 90032-8531}  } \date{\today}\maketitle
\begin{abstract}
We review various properties of the exceptional Euclidean Jordan algebra of degree three.  Euclidean Jordan algebras of degree three and their corresponding Freudenthal triple systems were recently shown to be intimately related to extremal black holes in $N=2$, $d=4$ homogeneous supergravities.  Using a novel type of eigenvalue problem with eigenmatrix solutions, we elucidate the rich matrix geometry underlying the exceptional $N=2$, $d=4$ homogeneous supergravity and explore the relations to extremal black holes.
\\\\
$Keywords:$ Jordan algebras, BPS black holes, U-duality.
\end{abstract}

\newpage
\tableofcontents
\section{Introduction}
\indent In recent papers \cite{1,2,3}, Gunaydin, Neitzke, Pioline and Waldron have proposed a program for counting microstates of $d=4$ BPS black holes in a class of $N=2$ supergravities whose scalar fields lie on a symmetric space $M_4=G_4/K_4$, called Maxwell-Einstein supergravity theories (MESGTs) or homogeneous supergravities \cite{1}.  These supergravity theories remain fairly mysterious, as most are not known to arise as low energy limits of a Calabi-Yau compactification of Type II string theory \cite{2}. The microstate counting program is motivated by the indications that three dimensional \emph{U}-duality groups $G_3$ yield spectrum generating symmetries for four dimensional BPS black holes \cite{1}.  Mathematically, such four dimensional BPS black holes were shown to be described by Freudenthal triple systems over Euclidean Jordan algebras of degree three \cite{5}.  Black holes with non homogeneous geometries have also been investigated by D'Auria, Ferrara and Trigiante in \cite{20}. \\
\indent Of the Euclidean Jordan algebras of degree three, there is an exceptional case, $\mathfrak{h}_3(\mathbb{O})$,  giving rise to an exceptional $N=2$, $d=4$ homogeneous supergravity.  In this paper, a central goal is to elucidate various properties of the exceptional Jordan algebra, which in turn sheds light on properties of the corresponding exceptional $N=2$, $d=4$ homogeneous supergravity.  
\section{$N=2$ Homogeneous Supergravities}
\subsection{Freudenthal Triple Systems}
In $N=2$, $d=4$ homogeneous supergravities there is a correspondence between the field strengths of the vector fields and their magnetic duals and elements of a Freudenthal triple system (FTS) $\mathcal{F}(\mathcal{J})$ over a Jordan algebra of degree three, $\mathcal{J}=\mathfrak{h}_3(\mathbb{A})$  \cite{5}.  The correspondence is explicitly:
\begin{equation}
\left(\begin{array}{ccc} F_{\mu\nu}^0 & & F_{\mu\nu}^i \\ & & \\ \tilde{F}_i^{\mu\nu} & & \tilde{F}_0^{\mu\nu} \end{array}\right) \Longleftrightarrow
 \left(\begin{array}{ccc} \alpha & & X \\ & & \\ Y & & \beta \end{array}\right)\in \mathcal{F}(\mathcal{J}),
\end{equation}
where $\alpha,\beta\in\mathbb{R}$ and $X,Y\in\mathfrak{h}_3(\mathbb{A})$, $\mathbb{A}=\mathbb{R},\mathbb{C},\mathbb{H},\mathbb{O}$.  $F_{\mu\nu}^i$ ($i=1,...,\hat{n}+1=n_V$) denote the field strengths of the vector fields from $d=5$ and $F_{\mu\nu}^0$ is the $d=4$ graviphoton field strength coming from the $d=5$ graviton \cite{5}.
\subsection{Extremal Black Holes}
\indent Using the correspondence, one can associate the entries of an FTS element with electric and magnetic charges $\{q_0,q_i,p^0,p^i\}\in\mathbb{R}^{2n_{V} + 2}$ of an $N=2$, $d=4$ extremal black hole via the relations \cite{2,3}:
\begin{equation}
\alpha=p^0 \quad \beta=q_0 \quad X=p^ie_i \quad Y=q_ie^i.
\end{equation}
Setting $p^I=(\alpha,X)$ and $q_I=(\beta,Y)$, the Bekenstein-Hawking entropy of such $d=4$ extremal black hole solutions is given by \cite{3}:
\begin{equation}
S_{BH}=\pi \sqrt{I_4(p^I,q_I)},
\end{equation}
where $I_4$ is the \emph{quartic invariant} of the FTS, preserved by the automorphism group $\textbf{Aut}(\mathcal{F}(\mathcal{J}))$.  The automorphism group is often written as $\textbf{Conf}(J)$, the four dimensional conformal \emph{U}-duality group of the $N=2$, $d=4$ extremal black hole \cite{3}. 

\section{Euclidean Jordan Algebras of Degree Three}

\indent A (real) Jordan algebra is a real vector space $\mathcal{A}$ equipped with a bilinear form (Jordan product) $(X,Y)\rightarrow X\circ Y$ satisfying $\forall X,Y\in\mathcal{A}$:
\begin{displaymath}
X\circ Y = Y\circ X
\end{displaymath}
\begin{displaymath}
X\circ (Y\circ X^2)=(X\circ Y)\circ X^2.
\end{displaymath}
The Euclidean Jordan algebras of degree three $\mathfrak{h}_3(\mathbb{A})$ are real Jordan algebras of $3\times 3$ Hermitian matrices over a division algebra $\mathbb{A}$ ($\mathbb{A}=\mathbb{R},\mathbb{C},\mathbb{H},\mathbb{O}$) satisfying $X^2+Y^2=0 \Longrightarrow X=Y=0\quad \forall X,Y\in\mathfrak{h}_3(\mathbb{A})$.

\subsection{The Exceptional Jordan Algebra}
\indent The Euclidean Jordan algebra of degree three over the octonions $\mathbb{O}$, called the exceptional Jordan algebra $\mathfrak{h}_3(\mathbb{O})$, is a real vector space of $3\times 3$ Hermitian matrices over $\mathbb{O}$ of the form:
\begin{equation}
X = \left(\begin{array}{ccc}a_1 & \varphi_1 & \overline{\varphi}_2 \\ \overline{\varphi}_1 & a_2 & \varphi_3 \\ \varphi_2 & \overline{\varphi}_3 & a_3 \end{array}\right) \qquad \qquad a_i \in \mathbb{R} \quad\varphi_j \in \mathbb{O}.
\end{equation}
The Jordan product in $\mathfrak{h}_3(\mathbb{O})$ is defined using ordinary matrix multiplication:
\begin{equation}
X\circ Y = \frac{1}{2}(XY+YX),
\end{equation}
where $XY$ is the product of $X$ and $Y$ using matrix multiplication.  It can be shown that the Jordan product in $\mathfrak{h}_3(\mathbb{O})$ satisfies the standard properties (where $X\circ X = XX=X^2$):
\begin{center}$X\circ Y = Y\circ X$,\end{center}
\begin{center}$X\circ (Y\circ X^2) = (X\circ Y)\circ X^2$. \end{center}
The trace $tr(X)$ is defined as usual for matrices.  The \textit{Freudenthal Product} $\ast: \mathfrak{h}_3(\mathbb{O}) \times \mathfrak{h}_3(\mathbb{O}) \rightarrow \mathfrak{h}_3(\mathbb{O})^{\ast}$ is defined in terms of the Jordan product by:
\begin{equation}
X\ast Y = X \circ Y - \frac{1}{2}(Y\textrm{tr}X+X\textrm{tr}Y)+\frac{1}{2}(\textrm{tr}X\textrm{tr}Y-\textrm{tr}(X\circ Y))I
\end{equation}
An important special case yields the \textit{quadratic adjoint map}:
\begin{equation}
X^\#=X\ast X = X^2 - (\textrm{tr}X)X + \sigma(X)I,
\end{equation}
a quadratic polynomial, where $\sigma(X)=\frac{1}{2}((\textrm{tr}X)^2-\textrm{tr}(X^2))=\textrm{tr}(X*X)$.\\
We can use the Freudenthal and Jordan product to define the \textit{cubic form}:
\begin{equation}
(X,Y,Z)=\textrm{tr}(X\circ (Y\ast Z)). 
\end{equation}
A special case of this cubic form is:
\begin{equation}
(X,X,X)=\textrm{tr}(X\circ(X \ast X))
\end{equation}
Through expansion by minors, the determinant of an element $X$ in $\mathfrak{h}_3(\mathbb{O})$ is calculated to be:
\begin{equation}
\textrm{det}(X)=a_1a_2a_3-a_1||\varphi_1||^2-a_2||\varphi_2||^2-a_3||\varphi_3||^3+2\textrm{Re}(\varphi_1\varphi_2\varphi_3).
\end{equation}
By direct computation, it is seen one can express the determinant as:
\begin{equation}
\textrm{det}(X)=\frac{1}{3}\textrm{tr}(X\circ(X \ast X))=N(X).
\end{equation}
where $N(X)$ denotes the \textit{cubic norm} of $X$.  This result also follows from the identity:
\begin{equation}
\textrm{det}(X)I=(X\circ(X \ast X)).
\end{equation}
A related identity which we will later find useful is:
\begin{equation}
X^{\#\#}=N(X)X.
\end{equation}
\subsection{Inner Product Space Properties}
An inner product over $\mathfrak{h}_3(\mathbb{O})$ as a vector space over $\mathbb{R}$ can be defined using the bilinear form:
\begin{equation}
\langle X,Y \rangle = \textrm{tr}(X\circ Y)\qquad X,Y\in\mathfrak{h}_3(\mathbb{O}).
\end{equation}
We see this is an inner product using the properties of trace and commutativity of $\mathfrak{h}_3(\mathbb{O})$:
\\\\
$\textbf{1.} \langle X,Y \rangle = \textrm{tr}(X\circ Y) = \textrm{tr}(Y\circ X) = \langle Y,X \rangle.$
\\
\\
$\textbf{2.} \langle X+Y,Z \rangle =\textrm{tr}((X+Y)\circ Z)=\textrm{tr}(X\circ Z)+\textrm{tr}(Y\circ Z)=\langle X,Z \rangle+\langle Y,Z \rangle.$
\\\\
$\textbf{3.} \langle cX,Y \rangle = \textrm{tr}(cX\circ Y)=c \langle X,Y \rangle,\quad c\in\mathbb{R}.$
\\\\
$\textbf{4.}$ Let $X = \left(\begin{array}{ccc}a_1 & \overline{\varphi}_1 & \varphi_2 \\ \varphi_1 & a_2 & \overline{\varphi}_3 \\ \overline{\varphi}_2 & \varphi_3 & a_3 \end{array}\right)$.\\
\indent  Then $\langle X,X \rangle =\textrm{tr}(X\circ X)=\textrm{tr}(X^2)$\\
\indent $=a_1^2+\overline{\varphi}_1\varphi_1+\varphi_2\overline{\varphi}_2+\varphi_1\overline{\varphi}_1+a_2^2+\overline{\varphi}_3\varphi_3+\overline{\varphi}_2\varphi_2+\varphi_3\overline{\varphi}_3+a_3^2\geq 0$, as $a_i^2$\\\indent and $\varphi_i\overline{\varphi}_i$ are non-negative real.
\\\\
$\textbf{5.}$ Suppose $\langle X,X \rangle =0$.\\
\indent This implies $a_1^2+\overline{\varphi}_1\varphi_1+\varphi_2\overline{\varphi}_2+\varphi_1\overline{\varphi}_1+a_2^2+\overline{\varphi}_3\varphi_3+\overline{\varphi}_2\varphi_2+\varphi_3\overline{\varphi}_3+a_3^2=0$.\\
\indent Since the $a_i^2$ and $\varphi_i\overline{\varphi}_i$ terms are non-negative real and sum to zero,\\
\indent it follows that all entries of $X$ are zero.  Hence, $X=0$.  The converse\\\indent is trivial, thus $\langle X,X \rangle=0$ if and only if $X=0$.\\\\
Using the inner product, one recovers a norm:
\begin{equation}
||X|| = \sqrt{ \langle X,X \rangle} =\sqrt{\textrm{tr}(X\circ X)}
\end{equation}
The metric induced by the norm then takes the form:
\begin{equation}
d(X,Y)=||X-Y||= \sqrt{ \langle X-Y,X-Y \rangle}.
\end{equation}
\textbf{Example 3.2.1}.\\
Using the inner product, the quartic invariant $I_4(p^I,q_I)$ is expressed as:
\begin{equation}
I_4(p^I,q_I)=-(\alpha\beta-\langle X,Y\rangle)^2-4(\alpha N(Y)+\beta N(X)-\langle X^{\#},Y^{\#}\rangle),
\end{equation}
where $\alpha,\beta\in\mathbb{R}$ and $X,Y\in\mathfrak{h}_3(\mathbb{O})$. In other literature, the terms $N(X)$ and $N(Y)$ are written as $I_3(X)$ and $I_3(Y)$, denoting the \emph{cubic invariant} of $\mathfrak{h}_3(\mathbb{O})$, preserved by the reduced structure group $\textbf{Str}_0(\mathfrak{h}_3(\mathbb{O}))=E_{6(-26)}$, the \emph{U}-duality group $G_5$ of the corresponding $N=2$, $d=5$ homogeneous supergravity with symmetric real manifold $\mathcal{M}_5=\frac{E_{6(-26)}}{F_4}$ \cite{4,5}.
\section{Eigenmatrices and Projective Space}
In \cite{7}, it was shown that the \textit{left} and \textit{right} eigenvalue problems for $X\in\mathfrak{h}_3(\mathbb{O})$ of the form ($v\in\mathbb{O}^3$):
\begin{center}
$Xv=\lambda v$
\end{center}
\begin{center}
$Xv=v \lambda$
\end{center}
admit quaternionic and octonionic eigenvalues.  This was unfortunate since for a Hermitian matrix one hopes for only real eigenvalues.  To remedy this Dray and Manogue proposed a new eigenvalue problem based on the Jordan product \cite{6}:
\begin{equation}
X\circ Q = \lambda Q\qquad X,Q\in\mathfrak{h}_3(\mathbb{O}).
\end{equation}
This new eigenvalue problem uses the \textit{left regular} representation of $\mathfrak{h}_3(\mathbb{O})$: the linear representation $L$ of $\mathfrak{h}_3(\mathbb{O})$ on $\mathfrak{h}_3(\mathbb{O})$ defined by $L(X)Q=X\circ Q$.  We will refer to this eigenvalue problem as the Jordan eigenvalue problem.  An \emph{eigenmatrix} of the Jordan eigenvalue problem is a non-zero $Q\in\mathfrak{h}_3(\mathbb{O})$ satisfying $X\circ Q = \lambda Q$. \\
\indent We are interested in solutions of the Jordan eigenvalue problem where the eigenmatrix satisfies $Q\ast Q=0$ for reasons what will be clear later.  Dray and Manogue formulated a modified characteristic equation to study such eigenmatrices \cite{6}. Using the determinant identity given earlier, their characteristic equation is:
\begin{equation}
\textrm{det}(X-\lambda I)I=(X - \lambda I)\circ ((X-\lambda I)\ast(X - \lambda I))=0.
\end{equation}
As was done in \cite{6}, we set
\begin{displaymath}
Q_{\lambda}=(X-\lambda I)\ast(X - \lambda I).
\end{displaymath}
(Note that $Q$ can be zero and if there is no ambigutity we omit the subscript.)  Substituting, we have
\begin{displaymath}
\textrm{det}(X-\lambda I)I=(X-\lambda I)\circ Q =0.
\end{displaymath}
Expanding, we see $Q$ is an eigenmatrix solution of the Jordan eigenvalue problem:
\begin{equation}
\textrm{det}(X-\lambda I)I=\frac{1}{2}(XQ-\lambda Q + QX-\lambda Q)=0\Longrightarrow (X\circ Q)=\lambda Q.
\end{equation}
Invoking identity (13) for $(X-\lambda I)$, we see $Q$ also satisfies:
\begin{equation}
Q\ast Q=(\textrm{det}(X-\lambda I))(X-\lambda I)=0.
\end{equation}
To verify the eigenvalues corresponding to non-zero eigenmatrices $Q$ are indeed real, we take the trace of both sides (20) yielding:
\begin{center}
$\textrm{tr}(X\circ Q)=\lambda\textrm{tr}(Q)$
\end{center}
If $Q$ is not traceless, it follows that $\lambda$ is real.  To confirm $Q$ is not traceless, we recall an eigenmatrix $Q$ is non-zero by hypothesis and that $Q\ast Q=0$, giving:
\begin{displaymath}
\textrm{tr}(Q\ast Q)=\frac{1}{2}((\textrm{tr}Q)^2-\textrm{tr}(Q^2))=0\Longrightarrow\textrm{tr}Q=\sqrt{\textrm{tr}(Q^2)}.
\end{displaymath}
From arguments used when defining the trace inner product, we have that $\textrm{tr}(Q^2)=0$ iff $Q=0$.  Since the eigenmatrix $Q$ is non-zero by assumption, $Q$ is not traceless and $\lambda$ is real.\\  
\indent Through direct computation it can be shown that \cite{6}:
\begin{displaymath}
(Q_1\circ X)\circ Q_2= Q_1\circ( X\circ Q_2)
\end{displaymath}
for eigenmatrices $Q_1$ and $Q_2$ of $X$ with eigenvalues $\lambda_1\neq\lambda_2$. As the eigenvalues are distinct, so are $Q_1$ and $Q_2$, and it follows that:
\begin{displaymath}
\lambda_1(Q_1\circ Q_2)=\lambda_2(Q_1\circ Q_2)\Longrightarrow Q_1\circ Q_2=0. 
\end{displaymath}
Using the trace inner product the previous result immediately yields:
\begin{displaymath}
\langle Q_1, Q_2 \rangle=0.
\end{displaymath}
Therefore, distinct eigenmatrices of the Jordan eigenvalue problem are mutually orthogonal under the trace inner product.  In the non-degenerate case ($\lambda_i \neq \lambda_j$), one may normalize the eigenmatrices and recover a basis for the span of the eigenspaces, a three dimensional subspace of the twenty-seven dimensional $\mathfrak{h}_3(\mathbb{O})$.  For brevity, we will refer to this three dimensional subspace as the stable space, $\textbf{stbl}(X)$, for $X\in\mathfrak{h}_3(\mathbb{O})$.  

\subsection{The Octonionic Projective Plane}
A projective \emph{n}-space over a field has points that are equivalence classes of (\emph{n}+1)-tuples satisfying $(x_0,x_1,...,x_n)=(\lambda x_0,\lambda x_1,...,\lambda x_n)$ for $\lambda\neq 0$.  This definition not only works for the fields $\mathbb{R}$ and $\mathbb{C}$, but also for the skew field $\mathbb{H}$ \cite{8}.  The definition, however, does not succeed in defining projective \emph{n}-space over $\mathbb{O}$ \cite{9}.  Fortunately, there do exist alternative definitions which succeed in defining a projective plane over $\mathbb{O}$ \cite{8,9,10}.  The most robust is that suggested by Sparling and Tillman in \cite{10}:\\\\
\textbf{Definition 4.1.1}\\
A point of $\mathbb{OP}^2$ is an equivalence class under (non-zero real) scaling of non-zero $J\in\mathfrak{h}_3(\mathbb{O})$ satisfying $J\ast J=0$.\\

\noindent With this definition, we see eigenmatrices $Q=(X-\lambda I)^{\#}$ of the Jordan eigenvalue problem, modulo real scaling, are points of $\mathbb{OP}^2$.\\

\noindent \textbf{Definition 4.1.2}\\
A line of $\mathbb{OP}^2$ is an equivalence class under (non-zero real) scaling of non-zero $L\in\mathfrak{h}_3(\mathbb{O})^{\ast}$ satisfying $L\ast L=0$.\\

\noindent \textbf{Definition 4.1.3}\\
A point $K$ lies on the line $L$ if and only if $\langle K, L\rangle=0$.\\

\noindent Under a duality map, which switches points and lines, the incidence relation is preserved \cite{8}.  To determine collinearity, we use the cubic form and decree: three points $P,Q,R\in \mathbb{OP}^2$ are collinear if and only if $(P,Q,R)=0$ \cite{8}.\\
\indent The group of collineations of the octonionic projective plane $\textbf{Coll}(\mathbb{OP}^2)$ is isomorphic to the reduced structure group $\textbf{Str}_0(\mathfrak{h}_3(\mathbb{O}))=E_{6(-26)}$ \cite{8,11}.  The isometry group $\textbf{Isom}(\mathbb{OP}^2)$ is isomorphic to the automorphism group $\textbf{Aut}(\mathfrak{h}_3(\mathbb{O}))=F_4$ \cite{8,9}.  Using these properties, the symmetric real manifold of $N=2$, $d=5$ homogeneous supergravity takes the more suggestive form:
\begin{equation}
\mathcal{M}_5=\frac{E_{6(-26)}}{F_4}=\frac{\textbf{Coll}(\mathbb{OP}^2)}{\textbf{Isom}(\mathbb{OP}^2)}.
\end{equation}
In the next section, we will further investigate the action of $E_{6(-26)}$ on $\mathbb{OP}^2$. 
\subsection{Attractors, Repellors and Saddle Points}
Generalizing the traditional three dimensional discrete dynamical system, we now describe a discrete dynamical system based on eigenmatrices.  Let $A$ be an element of the exceptional Jordan algebra with three orthogonal eigenmatrices $Q_1,Q_2,Q_3$ and corresponding eigenvalues $\lambda_1,\lambda_2,\lambda_3$.  If $Q_1,Q_2,Q_3$ are normalized, they form a basis for $\textbf{stbl}(A)$ (a projective basis for $\mathbb{OP}^2$).  Any initial vector $\textbf{X}_0$ in the stable space can be written as:
\begin{equation}
\textbf{X}_0=c_1Q_1+c_2Q_2+c_3Q_3.
\end{equation}
We act on such an initial vector with $A$ to acquire $\textbf{X}_1$:
\begin{equation}
\textbf{X}_1=A\circ \textbf{X}_0=c_1 (A\circ Q_1)+c_2 (A\circ Q_2)+c_3 (A \circ Q_3).
\end{equation}
\begin{displaymath}
\qquad\quad =c_1\lambda_1Q_1+c_2\lambda_2Q_2+c_3\lambda_3Q_3.
\end{displaymath}
We act again, identifying the next iterate $\textbf{X}_2$ as:
\begin{equation}
\textbf{X}_2=A\circ \textbf{X}_1=c_1\lambda_1 (A\circ Q_1)+c_2\lambda_2 (A\circ Q_2)+c_3\lambda_3 (A \circ Q_3).
\end{equation}
\begin{displaymath}
\qquad\quad =c_1(\lambda_1)^2Q_1+c_2(\lambda_2)^2Q_2+c_3(\lambda_3)^2Q_3.
\end{displaymath}
It is not difficult to see that in general
\begin{equation}
\textbf{X}_k=A\circ \textbf{X}_{k-1}=c_1(\lambda_1)^kQ_1+c_2(\lambda_2)^kQ_2+c_3(\lambda_3)^kQ_3.
\end{equation}
Thus, the long term behavior of the sequence $\textbf{X}_{0},\textbf{X}_{1},\textbf{X}_{2},...$ depends on the powers of the eigenvalues and not on the eigenmatrices.  This allows us to plot $\textbf{X}_{0},\textbf{X}_{1},\textbf{X}_{2},...$ as if they were points in $\mathbb{R}^3$.  The graph of such an iteration is called a \emph{trajectory} of the dynamical system.\\
\indent The magnitudes of the eigenvalues $\lambda_1,\lambda_2,\lambda_3$ determine the type of system associated with the matrix $A$.  For a given eigenvalue $\lambda_i$ we are interested in the magnitudes: $|\lambda_i| < 1$, $|\lambda_i| = 1$, and $|\lambda_i| > 1$.  This follows from the trajectory's dependence on powers of $\lambda$.  Collectively, the magnitudes of the eigenvalue set $\lambda_1,\lambda_2,\lambda_3$ determine if the origin is an \emph{attractor} ($|\lambda_i| < 1$), \emph{repellor} ($|\lambda_i| > 1$), \emph{saddle point} ($|\lambda_i| < 1$ and $|\lambda_j| > 1$), or \emph{steady state} ($|\lambda_i| = 1$).  Since the mapping $\textbf{X}_k\rightarrow \textbf{X}_{k+1}$ is linear, we have a linear dynamical system and the origin is the only possible attractor or repellor.\\
\indent As $E_{6(-26)}$ preserves the cubic norm, for $\phi\in E_{6(-26)}$ we have:
\begin{equation}
\textrm{det}(\phi(A-\lambda I))I=\textrm{det}(A-\lambda I)I.
\end{equation}
Written explicitly, this implies the following:
\begin{equation}
\phi(A-\lambda I)\circ(\phi(A-\lambda I)\ast\phi(A-\lambda I))=(A-\lambda I)\circ((A-\lambda I)\ast(A-\lambda I)).
\end{equation}
Thus, $\phi\in E_{6(-26)}$ map the eigenmatrices $Q_i=(A-\lambda_i I)\ast(A-\lambda_i I)$ of $A$ to the new eigenmatrices $Q^{'}_i=\phi(A-\lambda_i I)\ast\phi(A-\lambda_i I)$ of $A^{'}$ while preserving the eigenvalues.  Geometrically, the $E_{6(-26)}$ linear maps transform the basis for the dynamical system while preserving the attractor, repellor, or steady state property.
\subsection{Extremal Black Holes Revisited}
Using the techniques from earlier sections, we now explicitly calculate the eigenmatrices for a given element of $\mathfrak{h}_3(\mathbb{O})$ and find the entropy of the corresponding $N=2$, $d=4$ extremal black hole.  Let:
\begin{equation}
X = \left(\begin{array}{ccc}0 & \varphi_1 & 0 \\ \overline{\varphi}_1 & 0 & 0 \\ 0 & 0 & 0 \end{array}\right)\quad\varphi_1 \in \mathbb{O},\quad\varphi_1\overline{\varphi}_1=1.
\end{equation}
Computing the determinant of $(X-\lambda I)$ in the usual way, we find that $X$ has eigenvalues $\lambda_1=0$, $\lambda_2=-1$ and $\lambda_3=1$.  The corresponding eigenmatrices are $Q_1=X^{\#}$, $Q_2=(X+I)^{\#}$ and $Q_3=(X-I)^{\#}$, where $[Q_i,Q_j]=0$.  As $|\lambda_1|=0$, $|\lambda_2| = 1$ and $|\lambda_3| = 1$, the resulting dynamical system is a two dimensional steady state system, preserved by $E_{6(-26)}$ transformations.\\
\indent In \cite{12}, Krutelevich showed that by suitably acting with $\textbf{Aut}(\mathcal{F}(\mathcal{J}))$, one can bring an arbitrary element of an FTS to the simplified form:
\begin{equation}
 \left(\begin{array}{ccc} 1 & & Z \\ & & \\ 0 & & \beta \end{array}\right)\qquad\beta\in\mathbb{R},\quad Z\in\mathcal{J}=\mathfrak{h}_3(\mathbb{A}).
\end{equation}
It was shown in \cite{5} that $\textbf{Str}_0(\mathfrak{h}_3(\mathbb{A}))$ preserves such a simplified form.\\
\indent Using the simplified form, for arbitrary $\beta$, and noting that $N(X)=0$, we find the entropy of the corresponding $N=2$, $d=4$ extremal black hole to be:
\begin{equation}
S_{BH}=\pi \sqrt{-\beta^2}=\pi\beta i.
\end{equation}
In calculating the entropy we merely used the fact that $N(X)=0$, so the result holds for any element of $\mathfrak{h}_3(\mathbb{O})$ with vanishing determinant.  Since the entropy is imaginary, it may coincide
with the presence of a naked singularity \cite{13,14}.
\section{Conclusion}
We have shown how the linear space structure, regular representation and projective space structure of $\mathfrak{h}_3(\mathbb{O})$ elucidate the matrix geometry of the exceptional $N=2$, $d=4$ homogeneous supergravity.  In particular, eigenmatrices corresponding to a non-degenerate spectrum furnish a three dimensional subspace of $\mathfrak{h}_3(\mathbb{O})$ exhibiting attractor, repellor or steady state iterative properties, dependent on the magnitudes of the eigenvalues.  The physical interpretation of such eigenmatrices was not discussed; however, in the context of matrix models \cite{15,16,17,18,19}, they are perhaps special types of D0-branes characterizing exceptional $N=2$, $d=4$ extremal black holes.\\

\noindent \textit{A special thanks to professor Sultan Catto for inviting the author as an organizer at the 26th International Colloquium on Group Theoretical Methods in Physics, for which much of this work was prepared.}

\end{document}